\documentclass[aps]{revtex4}

\usepackage{amssymb}
\usepackage{graphicx}
\usepackage{subfigure}

\newif\ifpdf
\ifx\pdfoutput\undefined
\pdffalse 
\else
\pdfoutput=1 
\pdftrue
\fi
\ifpdf
\else
\fi

\begin{document}
\draft
\title{Towards the timely detection of toxicants}
\author{Massimiliano Ignaccolo$^{1,}$\footnote{Corrispondent Author. Email: mi0009@unt.edu (active till the end of September) \textbf{or} m\_ignaccolo@yahoo.it}}
\author{Paolo Grigolini$^{1,2,3}$}
\author{Guenter Gross$^{4}$}

\affiliation{$^{1}$Center for Nonlinear Science, University of North Texas,\\
P.O. Box 311427, Denton, Texas 76203-1427 }

\affiliation{$^{2}$Dipartimento di Fisica dell'Universit\`{a} di Pisa and
INFM, Via Buonarroti 2, 56127 Pisa, Italy }
\affiliation{$^{3}$Istituto di Biofisica del Consiglio Nazionale delle
Ricerche, Area della Ricerca di Pisa, \\ Via G. Moruzzi 1,
56124,
 Italy }
\affiliation{$^{4}$ University of North Texas Center for Network Neuroscience}
\date{\today}


\begin{abstract}
We address the problem of enhancing the sensitivity of biosensors to
the influence of toxicants, with an entropy method of analysis,
denoted as CASSANDRA, recently invented for the specific purpose of
studying non-stationary time series. We study the specific case where
the toxicant is tetrodotoxin. This is a very poisonous substance that
yields an abrupt drop of the rate of spike production at t approximatively
170 minutes when the concentration of toxicant is 4 nanomoles. The
CASSANDRA algorithm reveals the influence of toxicants thirty minutes
prior to the drop in rate at a concentration of toxicant equal to 2
nanomoles. We argue that the success of this method of analysis rests
on the adoption of a new perspective of complexity, interpreted as a
condition intermediate between the dynamic and the thermodynamic
state.Ê\end{abstract}
\maketitle

\section{Introduction}
One of the issues raised by September 11 is the timely detection of
toxicants.  Cell-based biosensors have proven to afford an efficient
way to monitor the presence in the environment of toxicants, pathogens
and other agents\cite{gross}.  In fact, upon increase of time the
function of the neural network (biosensor) is significantly affected
by the action of the toxicant, and the experimental results show that
the firing process undergoes significant changes under the presence of
toxicants. This important property of biosensors is shown by the
example of Fig. 1. At time $t_{1} = 30 $ minutes a given initial
concentration of toxicant, tetrodotoxin molecule, in this case, is
added to the compound, and it is steadily increased in time till at
time $t_{2} \approx 170 $ minutes, where by visual inspection
we see an abrupt change in the time dependence of the spike
rate for minutes, namely, the total number of spikes detected in one
minute, divided by the number of neurons monitored. A further increase
in the concentration of tetrodotoxin fully inihibits the spike
production at $t_{3} \approx 400$ minutes.

A theoretical challenge, for the Science of Complexity, is to detect
the presence of toxicant much earlier than the abrupt change in the
firing rate. This might become a question of death or life, if the
concentration of toxicants producing the abrupt change is very close
to the lethal concentration. An interesting issue is the following:
Let us consider the early portion of Fig. 1, in the time region
preceding, let us say, $t_{4} \approx 150$ minutes. Are there slight
dynamical changes that an ordinary visual inspection does not detect,
while a careful statistical analysis done with a fast computational
algorithm can? This is a problem of significant practical importance,
but at the same time is a theoretical challenge. In fact, if there are
slight changes, their intensity is expected to increase with
time. This means that the time series is not stationary, the
statistical analysis of a non-stationary time series being perceived
as a challenge. The title of the October 2002 Workshop, whose
Proceedings are published in this issue of Chaos, Solitons and
Fractals, containing this as well as many other papers originally
presented in this Workshop, is: \emph{Non-stationary Time Series: A
Theoretical, Computational and Practical Challenge}. The main purpose
of this paper is to address this challenging issue, in the special
case of biosensors, whose function is altered in time to an extent
that it is not easy to detect via mere visual inspection. We shall
prove that it is possible to address this issue with success, and that
the results obtained can be used to improve the biosensor technique,
with a significant progress towards the main goal of making a timely
detection of toxicants.


\begin{figure}[h]
\includegraphics[angle=-90,width=3.5in] {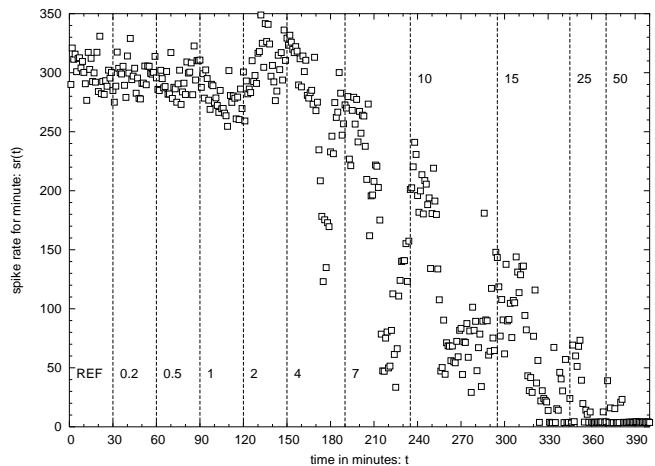}
\caption{The average spike rate for minute for the monitored potion of the network: $67$ neurons. 
The vertical lines signal the borders of states with a given concentration of tetradotoxin. The numbers between two consecutive vertical lines denote the concentration of tetrodotoxin in nano moles, characterizing the corresponding state. 
The label REF indicates the network state chosen as the reference for this experiment. In our case the reference state is 30 minutes of native (no toxicant added) activity where the average spike rate for minute is fairly stable.}
\label{figure1}
\end{figure}

\section{The CASSANDRA method}

The name CASSANDRA is the acronym for \emph{Complex Analysis of
Sequences via Scaling AND Randomness Assessment} \cite{cassandra}. The
authors of Ref. \cite{cassandra} proposed this algorithm for the
specific purpose of studying non-stationary time series. It is worth
reviewing the research work that led the authors of
Ref. \cite{cassandra} to the foundation of this method. This is
important because, as we shall see, this critical review has the
effect of proving that CASSANDRA is the natural consequence of
defining complexity as a condition intermediate between the dynamic
and the thermodynamic state.  This view of complexity, denoted as
Living State of Matter (LSM), has been advocated by the authors of the
work of Ref. \cite{gerardo}, proving the fruitful use of this
perspective for the foundation of a new theory for absorption and
emission of light in nano-materials. Actually, the authors of
Ref. \cite{gerardo} inspired their approach to non-living materials to
the perspective proposed by two professional biologists, whose ideas
are reviewed in a paper of these Proceedings \cite{marco}.  We hope
that the efficiency itself of the CASSANDRA algorithm, confirmed by
the results of the present paper, might afford further support to the
convenience of adopting the LSM perspective.

In principle, an attractive way to establish whether a time series is
totally random or not is given by the Kolmogorov-Sinai (KS) entropy
\cite{dorfman}. However, as widely discussed in an earlier publication
\cite{giulia}, there are problems with the computation estimate of the
KS entropy. Therefore, the authors of Ref. \cite{giulia} recommended
the recourse to two calculation techniques that are apparently
different but actually closely related the one to the other. The first
technique is the compression technique that rests on establishing if a
time series of a given length can be replaced by a much shorter
sequence, yielding the same information. From an intuitive point of
view, this compression is possible when the time series under study is
characterized to some extent by regularity. The second technique
discussed in Ref. \cite{giulia} is a method introduced for the first
time by the authors of Ref.\cite{nicola} and Ref.\cite{giacomo}.  This
method of analysis is based on converting the time series under study
into a diffusion process. This is done as follows. Let us denote by
$\{\xi_{i}\}$ the time series under study, with a total length $N$
that we assume to be very large. We define a discrete time $l$, which
is the size of a window that we move along the sequence. For any
window position $s$, $s = 1, 2, ...., N-l+1$, we select the values of
the sequence spanned by the window and we create the $s$-th diffusion
trajectory defined by
\begin{equation}
\label{generictrajectory}
x^{(s)} = \sum_{i}^{l} \xi(s+i).
\end{equation}
We consider this as the $s$-th random walker of a set of $N-l+1$
random walkers, moving for a time $l$, from the position $x= 0$ at
time $l=0$.  If the time series under study is totally random, the
random walkers are not correlated, and the resulting diffusion process
will be very close to the condition of ordinary diffusion. The method of Ref. \cite{nicola} and Ref. \cite{giacomo}
rests on evaluating the Shannon entropy of the resulting diffusion
process. For this reason the method was called Diffusion Entropy (DE)
method. According to a popular tenet of the Science of Complexity
\cite{mandelbrot}, complexity is related to the concept of scaling. In
the case of a diffusion process, with diffusion distribution density
$p(x,l)$, scaling is defined by
\begin{equation}
\label{scaling}
p(x,l) = \frac{1}{l^{\delta}} F(\frac{x}{l^{\delta}}).
\end{equation}
Complex systems are expected to generate a departure from the
condition of ordinary diffusion, where $\delta = 0.5$ and $F(y)$ is a
Gaussian function of $y$. This is very attractive and the DE method
yields the correct scaling coefficient even when the function $p(x,l)$
does not have a finite second moment, thereby invalidating the method
of scaling detection based on the variance evaluation. The DE method,
in the case when the property of Eq. (\ref{scaling}) applies, it is
easily proved to yield
\begin{equation}
\label{logarithmic}
S(l) = A + \delta \ln (l),
\end{equation}
where $A$ is a constant whose explicit expression is of no interest
here. It is evident that with this method the scaling parameter is
easily evaluated by plotting $S(l)$ in a linear-log representation.

However, in the case of a complex system, characterized by anomalous
scaling $\delta$, the time necessary for the system to make a
transition to the scaling regime, is extremely large and ideally
infinite, as proved by the authors of Ref. \cite{aging}. The work of
Ref.\cite{aging} suggests a very attractive way to define
complexity. Complexity is not so much represented by an anomalous
scaling, but rather by the transition regime necessary to reach the
scaling regime. The scaling regime is equivalent to the condition of
thermodynamic equilibrium. This is so because the scaling condition,
as the standard canonical equilibrium, is eventually realized moving
from any non-equilibrium (non-scaling) condition. However, it takes an
infinite, rather than a finite time, to realize the condition of
anomalous scaling. This would not be a problem, if the concentration
of tetrotodotoxin were kept constant. In practice, we guess that in a
suitably large time scale, of the order days, for instance, one would
be able to detect the anomalous scaling of the compound. The scaling
condition, in other words, is realized at a so large time scale, where
the memory effects produced by cooperation are lost, and the LSM
properties have been annihilated. In this scaling condition the
deviation of $\delta$ from $0.5$ and the deviation of $F$ from the
Gaussian shape, signal that something unusual occurred, when the
system was alive, namely, during the transition from the dynamic to
the thermodynamic condition. The anomalous property is the long time
spent by the system in a condition intermediate between dynamics and
thermodynamics. It is evident, therefore, that measuring the scaling
in practice is impossible when the dynamic rules are changed in a time
scale of the order of minutes. On the other hand, the anomalous
scaling is a signature of an anomalous transient condition, of significant
conceptual interest, which is even more important than the anomalous
scaling itself.

For all these reasons, the authors of Ref.\cite{cassandra} invented
the CASSANDRA algorithm. The name of this statistical method of
analysis seems to be very appropriate for the purpose of this paper,
since we shall be using it to predict much earlier the occurrence of
the toxicant-induced abrupt change of the firing activity. This is a
non-wished event belonging to the category of those that the
prophetess Cassandra would have predicted, and nobody would have
believed her. We shall prove that the CASSANDRA algorithm is as
accurate as the predictions of Cassandra prophetess, and we shall
prove that it deserves the reader's trust.  The CASSANDRA method works as
follows.  As pointed in Section 1, we study a time series that is not
stationary. This means that the rules change with time. However, let
us make the assumption that the change is slow enough that a portion
of the time series of length $L$ can be regarded as being a stationary
sequence. Then, we supplement the DE method with the introduction of a
big window of size $L$, which has to be considered as a sequence of
its own, and we move it along the sequence to study, for the main
purpose of assessing its local properties. The size of this window has
to be large enough as to make it possible to make a statistical
analysis. This means that for any window position $j$ we must span a
significant portion of the sequence $\{\xi\}$ under study, so as to
establish genuinely local properties. For each position of the big
window, using the procedure earlier described to create diffusion,
with small windows of size $l$, we evaluate the quantity

\begin{equation}
\label{cassandra}
C_{j} (\lambda)= \frac{\sum_{l=1}^{\lambda}[S_{j}(l) - S_{j}(1) - 0.5 \ln (l)]}{\lambda},
\end{equation}
where $S_{j}(l)$ and $S_{j}{(1)}$ denote the Shannon entropies of the
diffusion processes corresponding to small windows of size $l$ and
$1$, respectively, moving within the big window with position $j$. The
meaning of the prescriptionof Eq. (\ref{cassandra}) is transparent. We
compare the actual entropy change to the ideal change occurring with
an infinitely fast transition from dynamics to thermodynamics, this
corresponding, in fact, in the case of ordinary diffusion, to
the diffusion entropy increasing as $0.5 \ln(l)$.  Note that the
sequence $\{\xi\}$ will be built up in Section 4 with a prescription
that is the specific way we use to process the data of Fig. 1. This
prescription will establish a correspondence between the index $j$ and
the continuous time $t$, so that the index $j$ can be identified with
the continuous time $t$. Thus, for any value of the sequence $\{\xi\}$
we know the concentration of toxicant, which increases as we move from
the smaller to the larger values of the index $j$, denoting the
sequence site.

In conclusion, the validity of Eq. (\ref{cassandra}) as a measure of
the local deviation from the diffusion entropy increase that would be
generated by random fluctuations, rests on the mathematical inequality
\begin{equation}
\lambda << L << N.
\end{equation}
This is so because with the condition $L << N$, we can locate the big
window in different positions of the sequence $\{\xi\}$, corresponding
to no toxicants, to a small concentration of toxicants or to a large
concentration of toxicants, as we wish. The condition $\lambda << L$
makes it possible for us to use enough data so as to reach a
conclusion about the statistical property of the small region under
observation. We see from Eq.(\ref{cassandra}) that $C_{j}(\lambda)$
affords a significant information about the local process of
transition from dynamics to thermodynamics. If the small region under
observation is totally random, the transition to the scaling regime,
which in this case would be ordinary, is very quick, and the CASSANDRA
indicator yields a virtually vanishing value \cite{note}. If, on the
contrary, the small region under observation is compressible, in the
sense of Kolmogorov, as discussed in Section 2, the transition from
dynamics to thermodynamics (scaling regime) becomes much slower, and
the prescription of Eq. (\ref{cassandra}) yields values with a
significantly larger modulus, which, moreover, change with the changes
of the rules driving the dynamics of the process. This is the reason
of the sensitivity to a changing toxicant concentration, which is, in
fact, expected to change the dynamic rules of the process.

\section{Timely detection of toxicants}

This section is devoted to illustrating our method of analysis in
action on the analysis of the spike production of biosensors under the
influence of tetrodotoxin.  Tetrodotoxin is one of the most potent
molecules known. Once introduced, it selectively blocks the
voltage-sensitive sodium channels of excitable tissues. As a result,
tetrodotoxin inhibits or reduces the chances of an action potential to
be produced. Tetrodotoxin is complex in structure and contains a
imidazole ring. It is likely that this ring is the part of the
molecule that lodges in the channel leaving the rest of the molecule
blocking its outer mouth. It is therefore of crucial importance to
enhance the sensitivity of biosensors to this poisoning molecule. In
the first part of this section we describe how the experimental data
are processed, regardeless of whether this toxicant is acting or
not. Now we shall discuss this method of analysis in
action with the concentration of tetrodotoxin increasing in time.

\subsection{Random Walk representation of the neural netwotk activity}

The neural network (biosensor) is a composed by many neurons, but it
is possible to monitor only the activity of a part of it
\cite{gross}. For each neuron whose activity can be monitored we have
at our disposal, the time series representing the time at which the
neuron fires (time stamp series). Using the information about the
single neuron, we can build up a time series representing the temporal
evolution of the activity of the monitored neurons. This is done
dividing the time domain into intervals of time (bins) of duration
$\Delta$ (bin size) bigger or equal to the time resolution of the time
stamp series and counting the number spikes registered during each
interval of time.  Fig. 2 shows how $1$ minute of activity of the
monitored neurons is reproduced with the binning procedure described
above ($\Delta = 1/100$ seconds).


\begin{figure}[!h]
\includegraphics[angle=-90,width=3.5in] {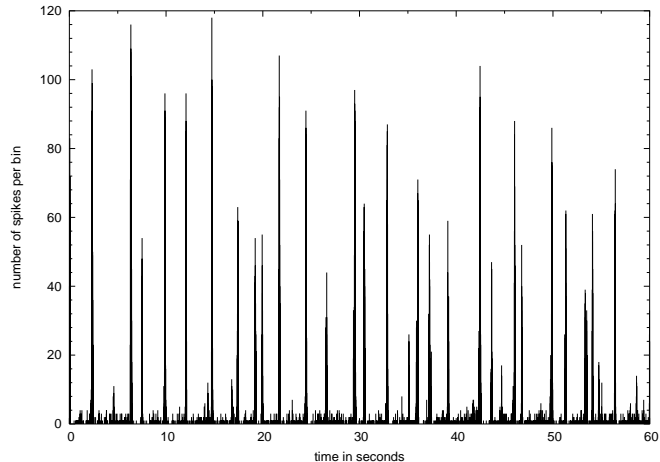}
\caption{A sample of our method of data processing. This is the number of spikes per bin, during the first minute of activity of the neural network. During this period of time no tetrodotoxin is present in the neural network.}
\label{figure2}
\end{figure}

The reason for this special way of data processing is to create
conditions corresponding to the complexity perspective established by
the work of Refs.\cite{giacomo, cassandra, giulia}, whose fundamental
prescription is in fact to turn the time series to analyze into a
continuous time random walk, and, this, in turn, into a diffusion
process. For any value of time of the abscissa axis, the random walker
makes a jump ahead by a quantity proportional to the number of spikes
per bin. This is a realization of the prescriptions of
Ref. \cite{giacomo}. Since we do not aim at establishing the scaling,
as in Ref.\cite{giacomo} for the reasons widely detailed in Section 1,
we do not care about the statistics of the jump intensities. In fact,
we shall be using CASSANDRA, and this algorithm will focus only on how
much the onset of resulting diffusion process departs from the ideally
instantaneous transition to ordinary diffusion.



\subsection{Experiment and Results}

As explained in Section 1 by the illustration of Fig. 1, the
experiment rests on adding the tetrodotoxin molecule to the compound,
with a concentration increasing with time. To make easier for the
reader to appreciate the benefit of using CASSANDRA, in Fig. 3 we
illustrate again the experimental results of Fig. 1, with a first
significant change occurring at about 170 minutes, the location of the
first dip.  Note that Fig. 3 refers to the case where the bin size has
the value $\Delta = 0.01 s$. We see that CASSANDRA produces its first
significant dip at 135 minutes, namely 35 minutes earlier! Changing
the bin size from $0.001 s$ to $0.1 s$ does not change the
result. This means that the changes observed by CASSANDRA are produced
by internal dynamics corresponding to that time scale.

\begin{figure}[h]
\includegraphics[angle=-90,width=3.5in] {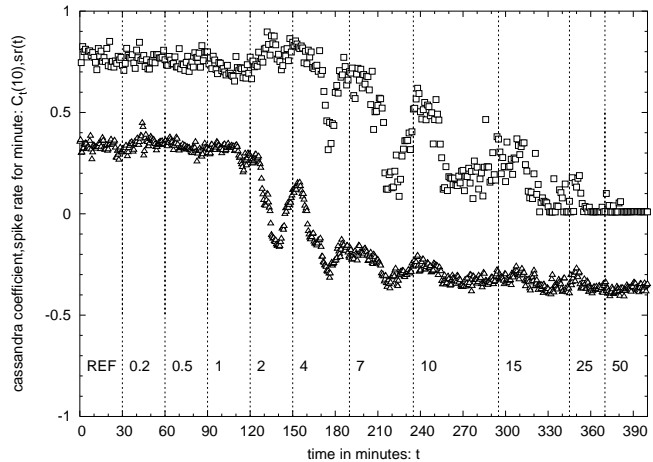}
\caption{The CASSANDRA coefficient (triangles) $C_{t}$ for $\lambda = 10$, as a function of time,  compared to a properly rescaled average spike rate for minute (squares). Note that the index $j$ of the prescription of Eq. (\ref{cassandra}) is replaced by the corresponding time $t$, which can be thought of as continuous. The meaning of the vertical lines is the sams as that of Fig. 1.}
\label{figure3}
\end{figure}

\begin{figure}[h]
\includegraphics[angle=-90,width=3.5in] {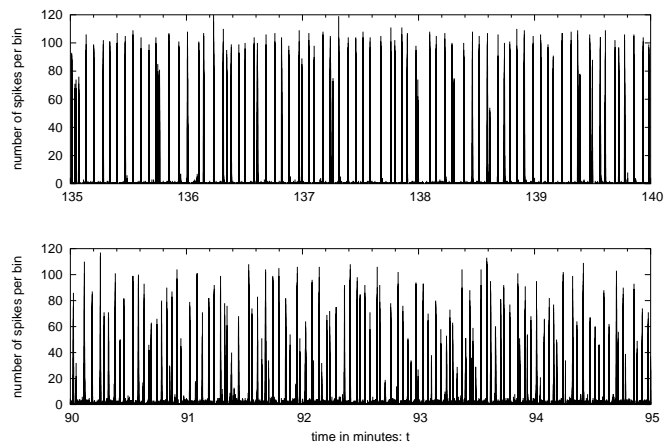}
\caption{Comparison of two different neural network activities. The bottom frame refers to  the time region betwen  90 and 95  minutes with the concentration  of 1 nanomole.  The top frame refers to the time region between 135 and 140 minutes, with a concentration of 2 nanomoles. }
\label{figure4}
\end{figure}

The importance of this result within the scenario of the bio-terrorism
threat is evident. However, here we want to focus our attention on the
use of CASSANDRA as a way to deepen our understanding of biosensor
dynamics. To shed more light into this issue, let us make some
comments regarding Fig. 4. This figure illustrates the spike production
occurring between $135$ and $140$ minutes and between $90$ and $95$
minutes. In the latter time interval the concentration is one
nano-mole and, at this small concentration, not even the CASSANDRA
algorithm succeeds in signaling the toxicant presence. In the
former, where the toxicant concentration is of 2 nano-moles, on the
contrary, CASSANDRA yields its first dip, thereby signaling the
toxicant presence, while the spike rate for minute remains fairly
stable (see Fig. 3).  The difference between the two kinds of activity
is made evident by these two frames. In the top frame the collective
burst events (high value of the number of spikes per bin) are more
regular in intensity (number of spikes) and, on average, keep one from
the next a time distance bigger than in the bottom frame. In other
words, there is a trade between time distance and intensity that has
the effect of leaving almost unchanged the rate of spike
production. These two different ways of realizing the same rate of
spikes per minutes is not overlooked by CASSANDRA, which in fact
signals the difference between the two conditions with the significant
dip at $t \approx 135$ minutes.



\section{Conclusion}
We are convinced that the surprising efficiency of CASSANDRA can be
given a convincing explanation, based on the LSM perspective
\cite{gerardo,marco}. In fact, the way we used to process the data
enhances the cooperative effect of the biosensors. These cooperative
effects are the main reason why the transition to scaling is delayed
with respect to ordinary Brownian motion. This is also the reason why,
as pointed out in Section 1, the adoption of a scaling detector would
not afford any significant information. In fact, the attainment of the
scaling condition is very slow, much slower than the change of
toxicant concentration as a function of time. The enhancement of the
collective properties corresponds to moving from the detailed
observation of a single channel, which might give the misleading
impression of dealing with dynamics, to the observation of collective
properties, related to the main idea of biosensors as an orchestra,
where the individual players are coordinated by the orchestra
director. Our LSM perspective refers to this cooperative condition,
which we do not have to confuse with the dynamic condition, taking
place at a much shorter time scale. The time scale between 0.1 s and
0.001 s is where the cooperative effects are significant. This is why
the transition to the scaling condition is very slow, with details
that are sensitive to the influence of toxicants. The toxicant
influences the cooperative behavior and consequently the regime of
transition to scaling (the thermodynamic regime). The CASSANDRA
algorithm, focusing on the observation of how the entropy increase
departs from the ideal behavior of a diffusion process with an
instantaneous transition to scaling, makes it possible to detect these
changes. The importance of this result within the threatening scenario
of bio-terrorism is evident.

P.G. and M.I. acknowledge support from ARO, through Grant DAAD19-02-0037.




\begin{references}
\bibitem{note} Actually, even in the case where the time series under study is random, it takes the diffusion process a finite time to reach the conventional scaling regime. This is so short, however, as to  fulfill  for moderately large values of $\lambda$ the condition that  $C_{j}(\lambda)$  vanishes if the local fluctuations are random.
\end{references}
\end{document}